\documentclass[epj]{svjour}
\usepackage{latexsym}
\usepackage[dvips,dvipdf]{graphicx}
\usepackage[dvips,usenames]{color}
\usepackage{epsf}
\newcommand{\cd}{\makebox[0.08cm]{$\cdot$}}

\begin{document}
\title{Solving Bethe-Salpeter equation for two fermions in Minkowski space}
\author{J. Carbonell\inst{1} \and V.A. Karmanov\inst{2}
}                     
%
%
\institute{Laboratoire de Physique Subatomique et Cosmologie,
CNRS/IN2P3,  53 av. des Martyrs,
38026 Grenoble, France \and Lebedev Physical Institute, Leninsky
Prospekt 53, 119991 Moscow, Russia}
\date{}
%
\abstract{The  method of solving the Bethe-Salpeter equation in
Minkowski space, developed previously for spinless particles
\cite{bs1}, is extended to a system of two fermions. The method is
based on the Nakanishi integral representation of the amplitude
and on  projecting the  equation on the light-front plane. The
singularities in the  projected  two-fermion kernel are
regularized  without modifying the original BS amplitudes. The
numerical solutions  for the $J=0$  bound state with the scalar,
pseudoscalar and massless vector  exchange kernels are found. The
stability of the scalar and positronium states without vertex form
factor is discussed. Binding energies are in close agreement with
the Euclidean results. Corresponding amplitudes in Minkowski space
are obtained.
\bigskip
} 
\maketitle

\section{Introduction}

Bethe-Salpeter  (BS) equation  for a relativistic bound system was
initially formulated in the Minkowski space \cite{BS}. It
determines the binding energy and the BS amplitude. However, in
practice, finding the solution in Minkowski space is made
difficult due its singular behaviour. The singularities are
integrable, but the   standard approaches for solving integral
equation fail. To circumvent this problem the BS equation is
usually transformed, by means of the Wick rotation,  into
Euclidean momentum space.

The binding energy provided by the Euclidean BS equation is the
same than the Minkowski one. For computing the binding energy it
is  enough to solve the Euclidean BS equation in the rest frame
$\vec{p}=0$. While the  rest frame solution is not enough to
obtain the electromagnetic form factors, since the initial and
final states cannot be simultaneously at rest. Contrary to the
Minkowski amplitude, the  dependence on the total momentum
$\vec{p}$ of the Euclidean amplitude is not given by a standard
boost, but must be found numerically by solving the Euclidean BS
equation for non-zero $\vec{p}$. This requires the rather
complicated calculations which were performed in
\cite{Maris1,Maris2}. Within this approach, the hadronic form
factors have been calculated, albeit with the aid of additional
assumptions.

However, the knowledge of the Euclidean amplitude in a moving
system is still not sufficient to calculate some observables, {\it
e.g.} electromagnetic form factors. The integral providing the
form factors contains singularities which are different from those
appearing in the BS equation. Their existence invalidates the Wick
rotation, to the zero component $k_0$, in the form factor integral
\cite{ckm_ejpa} and prevents from obtaining the exact form factors
in terms of the Euclidean BS amplitude alone. To avoid this
problem, the knowledge of the BS amplitude in Minkowski space is
mandatory. Thus, fifty years after its formulation, finding the BS
solutions in the Minkowski space is still a field of active
research. These solutions would pave the way to numerous
applications going from the hadronic electromagnetic form factors
of mesons  to the deuteron electrodisintegration amplitudes.

Some attempts have been recently made to obtain the Minkowski BS
amplitudes. The approach proposed in \cite{KW} is based on the
integral representation of the amplitudes and solutions have been
obtained for the ladder scalar case \cite{KW,KW1,SA_PRD67_2003} as
well as, under some simplifying ansatz, for the fermionic one
\cite{sauli}. Another approach \cite{bbmst} relies on a separable
approximation of the kernel which leads to analytic solutions.
Recent applications  to the $np$ system can be found in
\cite{bbpr}.

In a previous work \cite{bs1} we have proposed a new method to find
the BS amplitude in Minkowski space and applied it to the system
of two spinless  particles.

Our approach consists of two steps. In the first one,  the BS
amplitude is expressed  via the  Nakanishi integral representation
\cite{nakanishi1,nakanishi2}:

\begin{eqnarray}\label{bsint}
\Phi(k,p)&=&\int_{-1}^1\mbox{d}z'\int_0^{\infty}\mbox{d}\gamma'
\\&\times& \frac{g(\gamma',z')}{\left[  k^2+p\cdot k\; z'
+\frac{1}{4}M^2-m^2 -  \gamma' + i\epsilon\right]^3}. \nonumber
\end{eqnarray}

Notice that in this representation,  the dependence on the two
scalar arguments $k^2$ and $p\cd k$  of the BS amplitude is made
explicit by the integrand denominator and that the weight
Nakanishi  function $g(\gamma,z)$ is non-singular . By inserting
the amplitude (\ref{bsint}) into the BS equation on finds an
integral equation, still singular, for $g$.

In the second step, we apply to both sides of BS equation  an
integral transform -- light-front projection  \cite{bs1}  --
which eliminates singularities  of the BS amplitude.  It
consists in replacing $k\to k+\frac{\omega}{\omega \cd
p}\,\beta$ where $\omega$ is a light-cone four-vector
$\omega^2=0$, and integrating over $\beta$ in infinite limits.
We obtain in this way,  an equation for the non-singular
$g(\gamma,z)$. After solving it and substituting the solution in
eq. (\ref{bsint}), the BS amplitude in Minkowski space can be
easily computed.

As a first application  we have considered the spinless case
with ladder kernel \cite{bs1}. The binding energies were
compared  with the direct solution in the Euclidean space and
found to agree each other with high accuracy. By inserting the
computed weight function $g$ in (\ref{bsint}) and setting
$k_0=ik_4$ the result was found to coincide with the
corresponding Euclidean BS amplitude, found independently. The
method has been also successfully applied to the cross-ladder
kernel \cite{bs2} and to compute the electromagnetic form
factors \cite{ckm_ejpa}.

The main difference between our approach and  those followed in
\cite{KW,KW1,SA_PRD67_2003,sauli} is the use of the light-front
projection. This eliminates the singularities related to the BS
Minko\-w\-s\-ki amplitudes. The method is valid for any kernel
given by the irreducible Feynman graphs.

We present in this paper  the extension of our preceding work
\cite{bs1} to the two fermion system. In this case the Nakanishi
function $g$ is replaced by a set of $n$ functions $g_{i}$,
satisfying a system of coupled integral equations, with $n$
depending on  the total angular momentum  of the state ({\it e.g.}
$n$=4 for $J=0$, $n$=8 for  $J=1$). We will see that the direct
application to the fermionic kernels of the method used in the
spinless case, is however  marred with some numerical
difficulties. Although they could be overcome by a proper
treatment of the singularities, in this work we propose  an
alternative method   allowing to solve the BS equation for two
fermions in Minkowski space with the same degree of accuracy than
for the scalar case. The numerical applications will be limited to
the $J^{\pi}=0^+$ state.

The system of equations for the Nakanishi weight functions $g_i$
is derived in sect. \ref{deriv}  starting from the original BS
equation. In sect. \ref{Regularisation}  we develop a
regularization procedure for fermionic kernels. Numerical results
for the scalar, pseudoscalar and massless vector exchange
couplings are presented in sect. \ref{num}. Section \ref{concl}
contains concluding remarks. Some details of the calculations are
given in the appendices \ref{C_ij}, \ref{Derivation_Vd} and
\ref{Cb_ij}.

\section{Derivation of equation}\label{deriv}

The BS equation for the two fermions vertex function $\Gamma$
reads:
\begin{eqnarray}\label{psi1}
\Gamma(k_1,k_2,p)C &=& \int \frac{\mbox{d}^4k'}{(2\pi)^4} iK(
k_1,k_2,k'_1,k'_2 ) \cr &\times & \;\; \Gamma_1 S(k'_1)
\Gamma(k'_1,k'_2,p) C  S^t(k'_2) \Gamma_2^t
\end{eqnarray}
where $C=\gamma_2\gamma_0$ is the charge conjugation matrix, $S$
is the fermion propagator
$$
S(k'_{\alpha})=\frac{i(\hat{k'}_{\alpha}+m)}{{k'}^2_{\alpha}-m^2+i\epsilon}
\qquad \alpha=1,2
$$
$iK$ is the  interaction kernel and  $\Gamma_{\alpha}$ the
fermion-meson vertex. We denote by $\Gamma^t_{\alpha}$ its
transposed and $\hat{k}=k_{\nu}\gamma^{\nu}$. The charge
conjugation matrix $C$ appears here since we construct the vertex
function with two  fermions in the final state.

We have considered the following fermion ($\Psi,m$) - meson
($\phi,\mu$) interaction Lagrangians:

({\it i}) Scalar coupling
\begin{equation}\label{L_S}
{\cal L}_{int} ^{(s)}= g\, \bar{\Psi} \phi\, \Psi
\end{equation}
for which $\Gamma_{\alpha}=ig$

({\it ii}) Pseudoscalar coupling
\begin{equation}\label{L_Ps}
{\cal L}_{int} ^{(ps)}= ig\, \bar{\Psi}\gamma_5 \phi\, \Psi
\end{equation}
for which $\Gamma_{\alpha}=-g\gamma_5$.

 ({\it ii}) Massless vector exchange
\begin{equation}\label{L_V}
{\cal L}_{int} ^{(v)}= g\, \bar{\Psi} \gamma^{\mu}V^{\mu}\, \Psi
\end{equation}
with  $\Gamma_{\alpha}=ig\gamma^{\mu}$ and
$\Pi_{\mu\nu}=-i{g_{\mu\nu}/q^2}$  as vector propagator.

Each interaction vertex has been regularized with  a vertex
form factor $F(k-k')$ by  the replacement
\[ g \to gF(k-k') \]
and we have  chosen $F$  in the form:
\begin{equation}\label{ffN}
F(q)=\frac{\mu^2-\Lambda^2}{q^2-\Lambda^2+i\epsilon}.
\end{equation}

The BS amplitude $\Phi$ is defined in terms of the  vertex
function  $\Gamma$ by:
\[ \Phi(k_1,k_2,p)=S(k_1)\Gamma(k_1,k_2,p) S(-k_2) \]
Notice that we work with "unamputated" BS amplitude, which
includes the external propagators $S(k_1),S(-k_2)$.

Let us first consider the case of the scalar  coupling and  the
corresponding  ladder   kernel
 \[ K=\frac{1}{(k-k')^2-\mu^2+i\epsilon} \]
The BS equation for the amplitude $\Phi$ reads:
\begin{eqnarray} \label{bsf1}
\Phi(k,p)&=&\frac{i(m+\frac{1}{2}\hat{p}+\hat{k})}
{(\frac{1}{2}p+k)^2-m^2+i\epsilon}\; \int  \frac{\mbox{d}^4k'}{(2\pi)^4}\;\Phi(k',p) \\
&\times& \frac{(-ig^2)\,F^2(k-k')}{(k-k')^2-\mu^2+i\epsilon}\;\;
\frac{i(m-\frac{1}{2}\hat{p}+\hat{k})}{(\frac{1}{2}p-k)^2-m^2+i\epsilon},
\nonumber
\end{eqnarray}
where $p=k_1+k_2$, $k=(k_1-k_2)/2$,  $k'=(k'_1-k'_2)/2$.

In the case of  $J^{\pi}=0^+$ state, the  BS amplitude has the
following  general form:
\begin{equation}\label{bsf2}
\Phi(k,p)=S_1\phi_1+S_2\phi_2+S_3\phi_3+S_4\phi_4
\end{equation}
where $S_{i}$ are independent spin structures ($4\times 4$
matrices)  and $\phi_{i}$ are scalar functions of $k^2$ and
$p\cd k$.

The choice of $S_i$ is to some extent arbitrary.  To benefit
from useful orthogonality properties we have taken
\begin{eqnarray*}
S_1&=& \gamma_5 \\
S_2&=& \frac{1}{M}\hat{p}\,\gamma_5 \\
S_3&=& \frac{k\cd p}{M^3}\hat{p}\,\gamma_5-\frac{1}{M}\hat{k}\,\gamma_5,  \\
S_4&=& \frac{i}{M^2}\sigma_{\mu\nu}p_{\mu}k_{\nu}\,\gamma_5,
\end{eqnarray*}
where $ \sigma_{\mu\nu}=\frac{i}{2}(\gamma_{\mu}\gamma_{\nu}-
\gamma_{\nu}\gamma_{\mu})$.
The antisymmetry of the amplitude (\ref{bsf2}) with respect to  the permutation
$1\leftrightarrow 2$ implies for the scalar functions:
\begin{eqnarray}
\phi_{1,2,4}(k,p) &=& \phi_{1,2,4}(-k,p)\cr
\phi_{3}(k,p)       &=&-\phi_{3}(-k,p)  \label{phi_parity}
\end{eqnarray}

A decomposition similar to (\ref{bsf2}) was used in \cite{sauli}
to solve the BS equation for a quark-antiquark system but the
solution was approximated keeping only the first term
$S_1\phi_1$.

Substituting (\ref{bsf2}) in eq. (\ref{bsf1}),
multiplying it by $S_{i}$ and taking traces we get the following
system of equations for the invariant functions $\phi_{i}$:
\begin{small}
\begin{eqnarray}\label{bsf4}
&&\phi_a(k,p)=\frac{i} {[(\frac{1}{2}p+k)^2-m^2+i\epsilon]}
\frac{i}{[(\frac{1}{2}p-k)^2-m^2+i\epsilon]} \nonumber\\
\lefteqn{\int \frac{\mbox{d}^4k'}{(2\pi)^4}
\frac{(-ig^2)F^2(k-k')}{(k-k')^2-\mu^2+i\epsilon}
\sum_{a'=1}^4c_{aa'}(k,k',p)\phi_{a'}(k',p)}
\end{eqnarray}
\end{small}

For scalar and pseudoscalar couplings, the coefficients $c_{ij}$
are given by:
\begin{small}
\begin{equation}\label{cij}
c_{ij}=\frac{1}{N_i}Tr\left[S_i
\left(\frac{\hat{p}}{2}+\hat{k}+m\right)\Gamma_{1}
S'_j\bar{\Gamma}_{2}\left(\frac{\hat{p}}{2}-\hat{k}-m\right)\right]
\end{equation}
\end{small}
where $N_i=Tr[S_i^2]$ are the normalization factors, $S'_j$ is
obtained from $S_j$ by the replacement $k\to k'$,
$\bar{\Gamma}_{\alpha} = U_c\Gamma^t_{\alpha}
U_c=\Gamma_{\alpha} $ with $\Gamma_{\alpha} $  given after eqs.
(\ref{L_S}) and (\ref{L_Ps}).

For the fermion-fermon vector massless coupling, the $c_{ij}$
coefficients are also given by eq. (\ref{cij}) with the
replacement $\Gamma_1 \to g\gamma_{\nu}$ and $\bar{\Gamma}_2 \to
-g\gamma^{\nu}$ which must be contracted. For the positronium
case (fermion-antifermion), they are multiplied by an extra factor $-1$.

The explicit expression of  $c_{ij}$  for a $J^{\pi}=0^+$ state
and the scalar coupling (\ref{L_S}) are
\begin{eqnarray*}
c_{11} &=&m^2+\frac{1}{4}M^2-k^2\\
c_{12} &=&mM \\
c_{13} &=& c_{31}=0 \\
c_{14} &=&-b' M^2\\
c_{21} &=& mM\\
c_{22} &=&m^2+\frac{1}{4}M^2+k^2 -\frac{2(k\cd p)^2}{M^2} \\
c_{23} &=&-2b'(p\cd k)  \\
c_{24}&=&-2b' m M  \\
c_{32} &=& 2(p\cd k)  \\
c_{33} &=& \frac{b'}{b}\left(m^2-\frac{1}{4}M^2
+2\frac{(p\cd k)^2}{M^2}-k^2\right) \\
c_{34}&=&2\frac{b'}{b}  \frac{m}{M}(p\cd k) \\
c_{41} &=&M^2\\
c_{42} &=&2 m M \\
c_{43} &=& 2\frac{b'}{b} \frac{m}{M} (p\cd k)\\
c_{44} &=& -\frac{b'}{b} \left(\frac{1}{4}M^2-m^2-k^2\right)
\end{eqnarray*}
where
\begin{eqnarray*}
b  &=& \frac{1}{M^4}\Bigl[(p\cd k)^2-M^2k^2\Bigr]  \\
b' &=& \frac{1}{M^4}\Bigl[(p\cd k)(p\cd k')-M^2(k\cd k')\Bigr]
\end{eqnarray*}

The  coefficients $c_{ij}$  for the pseudoscalar exchange
(\ref{L_Ps})   are simply obtained by changing the sign of the
$j=1$ and $j=4$ matrix elements in the scalar ones:
\begin{equation}\label{C_Ps}
c^{PS}_{ij=1,4}= - c^{S}_{ij=1,4}   \qquad , \forall i
\end{equation}

For the massless vector exchange (\ref{L_V})  the coefficients are given by
\begin{equation}\label{C_V}
Ê  c^{V}_{ij}=   \xi _{ij}\;  c^S_{ij},   \qquad
\xi  =\pmatrix{ 4&-2&0&0\cr 4&-2&-2&0  \cr 0
&-2&-2&0 \cr 4 &-2 &-2 &0}
\end{equation}

\bigskip
The equation (\ref{bsf4}) is the Minkowski space BS equation for
two  fermions we aim to solve. As in the scalar case, the  BS
amplitudes $\phi_i$ are singular  and a direct solution of
(\ref{bsf4})  is not suitable even for the simplest kernels.

To overcome this difficulty, the first idea is to represent each
of the  BS components $\phi_i(k,p)$ by means of the Nakanishi
integral
\begin{eqnarray}\label{Nakanishi_phi}
\phi_i(k,p)&=&\int_{-1}^1\mbox{d}z'\int_0^{\infty}\mbox{d}\gamma'
\\&\times& \frac{g_i(\gamma',z')}{\left[  k^2+p\cdot k\; z'
+\frac{1}{4}M^2-m^2 -  \gamma' + i\epsilon\right]^3}. \nonumber
\end{eqnarray}
and apply the light-front projection to the  set of coupled
equations for the corresponding weight functions $g_i(\gamma,z)$.
As mentioned in the Introduction, this projection, which is an
essential ingredient of our previous works \cite{bs1,bs2},
consists in replacing $k\to k+ {\omega\over \omega\cdot p}\beta$
in eq. (\ref{bsf4})  and integrating over $\beta$ in all the real
domain. This integration is quite similar to the case of the
spinless particles explained in detail in \cite{bs1}.

We obtain in this way a set of coupled two-dimensional integral
equations which can be written in the general form valid for all
types of couplings and states:
\begin{eqnarray}\label{BSMF}
\int_{0}^{\infty} \mbox{d}\gamma'  \frac{1} {\left[D_0(\gamma,z)
+\gamma'\right]^2}   \; g_{i}(\gamma',z) =&& \cr \sum_{j}
\int_{0}^{\infty} \mbox{d}\gamma'\int_{-1}^{+1} \mbox{d}z'  \;
V_{ij}(\gamma,z ,\gamma',z') \;g_{j}(\gamma',z')&&
\end{eqnarray}
where
\[ D_0(\gamma,z) =\gamma +m^2z^2+(1-z^2)\kappa^2  \]
and
\[  \kappa^2 = m^2-{M^2\over4} \]
The left hand side of (\ref{BSMF}) is the same for all  the
amplitudes and coincides with the scalar case \cite{bs1}.

The interaction kernel $V_{ij}$ can be written in the form
\begin{equation}\label{VMF}
V_{ij}(\gamma,z,\gamma',z') = \left\{
\begin{array}{rl}
                W_{ij}(\gamma,z,\gamma',z') &{\rm if} \   z'\leq z \\
\sigma_{ij}\; W_{ij}(\gamma,-z,\gamma',-z') &{\rm if} \ z'> z
\end{array} \right.
\end{equation}
with
\begin{equation}\label{epsilon}
\sigma =\pmatrix{+1&+1&-1&+1\cr +1&+1&-1&+1\cr-1&-1&+1&-1\cr+1&+1&-1&+1}
 \end{equation}
and
\begin{eqnarray}\label{W}
 W_{ij}(\gamma,z,\gamma',z')  = {\alpha m^2\over2\pi} \frac{(1-z)^2}{D_0} &&\cr
  \int_0^1 Z_{\Lambda}(\gamma,z,\gamma',z';v) \;  C_{ij}(\gamma,z;v)\;
  {v^2\over D^2} \mbox{d}v  &&
\end{eqnarray}
We use here  $\alpha={g^2\over 4\pi}$,  whereas the denominator
\begin{eqnarray*}
D(\gamma,z,\gamma',z', v  )&=&   v(1-v)  (1-z') \gamma \cr
 &+&   v   m^2 \;[  (1-v)   (1-z')z^2 + v{z'}^2(1-z) \;]\cr
 &+&   v   \kappa^2 (1-z)(1-z') \;[  1+z-v(z-z') \;] \cr
 &+& (1-z) [ (1-v) \mu^2 +  v  \gamma']
\end{eqnarray*}
is  the  same than in our previous work \cite{bs1}.

The function
\begin{eqnarray*}
Z_{\Lambda} &=&    (1-v)^2(1-z)^2\\
& \times&  \frac{\left[3
D+(1-v)(1-z)(\Lambda^2-\mu^2)\right](\Lambda^2-\mu^2)^2}{
[D+(1-v)(1-z)(\Lambda^2-\mu^2)]^3}
\end{eqnarray*}
contains the dependence on the vertex form factor parameter $\Lambda$ with
\[  \lim_{\Lambda\to\infty} Z_{\Lambda}= 1\]
The matrix coefficients  $C_{ij}$   couple  the different spin
components.  They are given in appendix \ref{C_ij} for the
$J^{\pi}=0^+$ states. Notice that setting $C_{ij}=\delta_{ij}$ and
$Z_{\Lambda}=1$ the set of eqs. (\ref{BSMF}) decouples and each of
them is identical to the spinless case one given in \cite{bs1}.

\begin{figure}[ht!]
\vspace{.5cm}
\begin{center}\includegraphics[width=8cm]{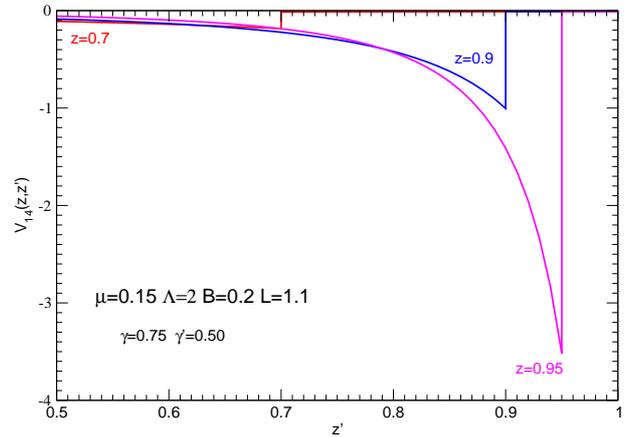}
\end{center}
\caption{The kernel matrix element $V_{14}(\gamma,z;\gamma',z')$,
eq. (\ref{VMF}), for $z=0.7,0.9,0.95$ as a function of $z'$ and
fixed values of $\gamma,\gamma'$. The discontinuity is finite at a
fixed value of $z$ but diverges when $z\to1$.} \label{V14}
\end{figure}

\bigskip
It turns out  that, in contrast to the scalar case, most of the
kernel  matrix elements $V_{ij}$  are discontinuous at $z'=z$. In
some cases -- like {\it e.g.}     $V_{14}$ displayed in fig.
\ref{V14} --   the value of the discontinuity, although being
finite at fixed value of $z$, diverges when $z\to\pm1$. This
creates some numerical difficulties when computing the solutions
$g_i(\gamma,z)$ in the vicinity of $z=\pm1$. They can be in
principle solved by properly taking into account the particular
type of divergence. However the latter may depend on the
particular matrix element, on the type of coupling, the quantum
number of the state and other details of the calculation. We
propose in next section an alternative, and efficient way to
overcome this difficulty.

\section{Regularized kernels}\label{Regularisation}

To deal with more regular kernels, the original BS equation
(\ref{bsf4}) is first multiplied on  both sides of by the factor
\begin{equation}\label{eta}
\eta(k,p)=\frac{(m^2-L^2)}{\left[(\frac{p}{2}+k)^2-L^2+i\epsilon\right]}
\frac{(m^2-L^2)}{\left[(\frac{p}{2}-k)^2-L^2+i\epsilon\right]}
\end{equation}
This factor has the form of a product of two scalar propagators with
mass $L$. It plays the role of form factor suppressing the  high
off-mass shell values of the constituent four-momenta
$k^2_{1,2}=(\frac{p}{2}\pm k)^2$ and tends to 1  when $L\to
\infty$.

The light-front projection and Nakanishi transform are then applied to
the equation
\begin{eqnarray}\label{bsf4p}
&&\eta(k,p)\;\phi_i(k,p)= \\
&& \frac{\eta(k,p)} {[(\frac{p}{2}+k)^2-m^2+i\epsilon]
[(\frac{p}{2}-k)^2-m^2+i\epsilon]} \nonumber\\
&\times& \int \frac{\mbox{d}^4k'}{(2\pi)^4} \frac{i
g^2\,F^2(k-k')}{(k-k')^2-\mu^2+i\epsilon}
\sum_{j=1}^4c_{ij}(k,k',p)\phi_j(k',p), \nonumber
\end{eqnarray}
Since $\eta(k,p)\neq 0$, the equation thus obtained is strictly
equivalent to (\ref{bsf4}). We will see however that, due to the
presence of the $\eta$ factor, the light front projection
modifies the resulting kernels which become less singular
functions.

\begin{figure}[h!]
\vspace{.5cm}

\begin{center}\includegraphics[width=8.cm]{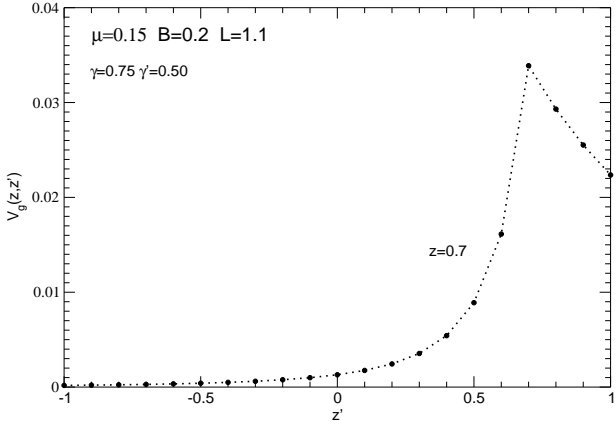}\end{center}
\caption{The left hand side kernel
$V^g(\gamma=0.75,z=0.7,\gamma=0.5,z')$, eq. (\ref{Vg}),  as a
function of $z'$ for L=1.1\,m}\label{fig_Vg}
\end{figure}

The technical details of the light-front projection are similar
to those  given in ref. \cite{bs1}. The differences, due to the
factor $\eta(k,p)$, are explained in Appendix
\ref{Derivation_Vd}.  The new set of equations has the following
form:
\begin{eqnarray}\label{eq0f}
&&\int_0^{\infty}\mbox{d}\gamma'\int_{-1}^1\mbox{d}z' \;
V^g(\gamma,z;\gamma',z')\;
g_i(\gamma',z') = \nonumber \\
&&\sum_{j}\int_0^{\infty}\mbox{d}\gamma'\int_{-1}^{1}\mbox{d}z'
\;V^d_{ij}(\gamma,z;\gamma',z') g_j(\gamma',z')
\end{eqnarray}

The kernels $V^g$ and $V^d_{ij}$ depend now on the parameter
$L$. Closer is $L$ to  $m$, smoother is the kernel  and more
stable are the  numerical solutions.  However  the    weight
functions $g_i(\gamma,z)$ as well as  binding energies provided
by  (\ref{eq0f})  are independent of $L$.

Notice that, in contrast to (\ref{BSMF}), the left hand side of
eq. (\ref{eq0f}) is also a two-dimensional integral. The
corresponding  kernel $V^g$  is the same for all the components
and has the form:
\begin{equation} \label{Vg}
V^g(\gamma,z;\gamma',z')=\left\{
\begin{array}{ll}
W^g(\gamma,z;\gamma',z')      &\;\;\mbox{if $ z'\le z$}\\
W^g(\gamma,-z;\gamma',-  z') &\;\; \mbox{if $ z'> z$}
\end{array}\right.
\end{equation}
with
\begin{equation}\label{f}
W^g(\gamma,z;\gamma',z')=\frac{(L^2-m^2)}{H^3}
\end{equation}
and
\begin{eqnarray*}
H &=& \gamma\frac{(1-z')}{(1-z)}+\gamma' +(1-z')(1+z)\kappa^2 \\
    &+& \Bigl(z'-z(1-z')\Bigr)m^2 +\frac{(z-z')}{(1-z)}L^2
\end{eqnarray*}

To avoid spurious singularities in (\ref{f}) due to the $\eta$
factor   (\ref{eta}), $L^2$ must be larger than
$\frac{M^2}{2}-m^2$, what  is fulfilled for $L>m$. In practical
calculations we have taken $m=1$ and $L=1.1$.

\begin{figure}[h!]
 \vspace{.5cm}
\begin{center}\includegraphics[width=8cm]{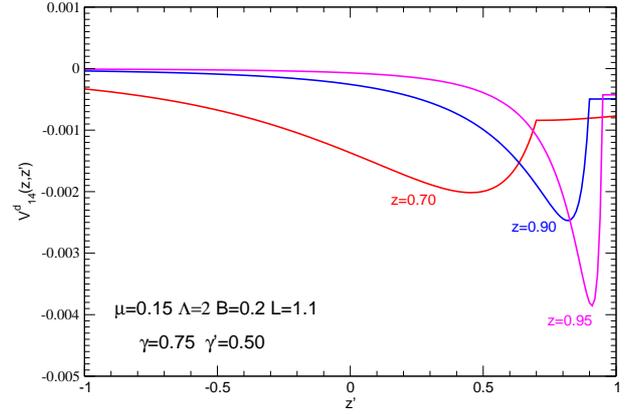}
\end{center}
\caption{Regularized right hand side  kernel
$V^d_{14}(\gamma,z;\gamma',z')$, eq. (\ref{Vij}),  v.s. fixed
values of $z=0.7,\;0.9,\;0.95$ and  $L=1.1\,m$. Compare with fig.
\ref{V14}.} \label{V14_d}
\end{figure}

The kernel  $V_g$  is finite and vanishes for $z=\pm1$. For a
fixed values  of $\gamma,z$ and $\gamma'$, $V_g$ is a continuous
function of  $z'$ with a discontinuous derivative at $z'=z$. It is
represented in fig. \ref{fig_Vg} for  $z=0.7$.

\bigskip
The right hand side  kernel $V^d_{ij}$ is  given by:
\begin{equation}\label{Vij}
V^d_{ij}(\gamma,z;\gamma',z')=
 \left\{\begin{array}{ll}
\phantom{ \sigma_{ij}}  W^d_{ij}(\gamma,z;\gamma',z')   &\; \mbox{if $z'<z$}  \\
\sigma_{ij}   W^d_{ij}(\gamma,-z;\gamma',-z') &\;\mbox{if
$z'>z$}
 \end{array}
 \right.
\end{equation}
$W^d_{ij}$  is represented as a sum of two terms:
\[W^d_{ij}= W^a_{ij}+W^b_{ij} \]
which can be written in the same form than (\ref{W})
\begin{eqnarray}\label{W_ab}
 W^{\nu}_{ij}(\gamma,z,\gamma',z')  = {\alpha m^2\over2\pi}
 \frac{(1-z)^2}{D_0}  F^{\nu}_L(\gamma,z)&&\cr
 \times  \int_0^1 Z^{\nu}_{\Lambda}(\gamma,z,\gamma',z';v) \;
  C^{\nu}_{ij}(\gamma,z;v)\; {v^2\over D_{\nu}^2} \mbox{d}v   &&
\end{eqnarray}

For the first term ($\nu=a$) one has
\begin{eqnarray*}
F^a_L  &=& \frac{(1-z) \;[ D_0 + (L^2-m^2)]}{2D_0+ (1-z)(L^2-m^2)}  \\
Z^{a}_{\Lambda} &=& Z_{\Lambda} \\
C^{a}_{ij} &=& C_{ij}\\
D_a   &=& D
\end{eqnarray*}
where $Z_{\Lambda} ,C_{ij}$ and $D$   are the same than in eq.
(\ref{W}).

For the second term ($\nu=b$) one has
\begin{eqnarray*}
F^b_L  &=&  -\frac{D_0(1-z)}{2D_0 + (1+z)(L^2-m^2)}   \\
Z^b_{\Lambda} &=&    (1-v)^2(1-z)^2\\
 & \times&  \frac{\left[3 D_b+(1-v)(1-z)(\Lambda^2-\mu^2)\right]
 (\Lambda^2-\mu^2)^2}{ [D_b+(1-v)(1-z)(\Lambda^2-\mu^2)]^3}  \\
D_b   &=& D_a  + v(1-v)(z-z')(L^2-m^2)
\end{eqnarray*}
with the coefficients $C^b_{ij}$ given in Appendix \ref{Cb_ij}.

Notice that, like $Z_{\Lambda}$,  $Z^b_{\Lambda}\to 1$ when
$\Lambda\to\infty$. Notice also that $F^a_L\to1$ and $F^b_L\to0$
in the limit $L\to\infty$. One has consequently $W^a=W$, $W^b=0$
and $V^d=V$  given in (\ref{VMF}). Furthermore the peak at $z'=z$
in the left kernel $V^g$ displayed in fig. \ref{fig_Vg} tends to a
delta function $\delta(z'-z)$  with a coefficient which reproduces
the left-hand side of eq. (\ref{BSMF}).

For a finite value of $L$, both systems of equations are also
strictly equivalent to each other but the $z'$-dependence of the
regularized kernels is much more smooth and therefore better
adapted  for obtaining accurate numerical solutions. In fig.
\ref{V14_d} we plotted the regularized kernel $V^d_{14}$ as a
function of $z'$ for the same arguments $\gamma,z,\gamma'$ and
parameters than in  fig. \ref{V14}, where it was calculated
without the $\eta(k,p)$ factor. As one can see, the kernel is now
a continuous function of $z'$. A discontinuity of the first
derivative however remains at $z'=z$.

Some additional remarks concerning the regularization  procedure
presented in this section are in order:

{\it (i)}   The most singular kernel is $V_{23}$.   It is the
only matrix element that after regularization by means of the
$\eta(k,p)$ factor remains discontinuous at $z'=z$ . The degree
of the discontinuity in the limit $z\to \pm1$ is however reduced
and does no longer spoil the numerical accuracy.

{\it (ii)}  This procedure can be improved by choosing  for
$\eta(k,p)$ a more strongly decreasing function of the arguments
$\left(\frac{p}{2}\pm k\right)^2$. The simple replacement
$\eta(L)\to \eta(L_1)\eta(L_2)$ would reduce the degree of
singularity of  the remaining singular matrix elements in the
limit $z\to\pm1$. We have not found  any need for that in the
present calculations.

{\it (iii)}   We would like to emphasize again that despite the
fact that the non-regularized and regularized kernels are very
different from each other (compare {\it e.g.} the figs. \ref{V14}
and \ref{V14_d}) and that the regularized ones strongly depends on
the value of $L$, they provide -- up to numerical errors -- the
same binding energies and  weight functions $g_i(\gamma,z)$. We
construct in this way a family of equivalent kernels.

\begin{table}[ht!]
\begin{center}
\caption{Coupling constant $g^2$ as a function of  binding energy
$B$   for  the $J=0$ state with scalar (S), pseudoscalar (PS) and
massless vector (positronium) exchange kernels. The vertex form
factor is $\Lambda=2$ and the parameter of the $\eta$ factor
$L=1.1$.} \label{tab_B_S_Ps}
\begin{tabular}{c|cc |cc |cc |c}
           &  S                 &                         & PS  &     & positronium\\ \hline
$\mu$&  0.15           & 0.50        & 0.15      & 0.50 &0.0           \\ \hline
$B$    & $g^2$         & $g^2$     & $g^2$  & $g^2$  & $g^2$              \\
0.01   &   7.813        &  25.23     &   224.8  &  422.3         &     3.265        \\
0.02   &   10.05        &  29.49     &   232.9  &  430.1         &     4.910 \\
0.03   &   11.95        &  33.01     &   238.5  & 435.8          &     6.263 \\
0.04   &   13.69        &  36.19     &   243.1  & 440.4          &    7.457 \\
0.05   &   15.35        &  39.19     &   247.0  &  444.3         &    8.548 \\
0.10   &   23.12        &  52.82     &   262.1  &   459.9        &  13.15  \\
0.20   &   38.32        &  78.25     &   282.9  &   480.7        &   20.43\\
0.30   &   54.20        & 103.8      &   298.6  &   497.4        &  26.50    \\
0.40   &   71.07        & 130.7      &    311.8 &     515.2      &   31.84 \\
0.50  &   86.95         & 157.4      &    323.1 &     525.9      &   36.62 \\
\end{tabular}
\end{center}
\end{table}

\section{Numerical results}\label{num}

The solutions of eq. (\ref{eq0f}) have been obtained using the
same  techniques than in ref \cite{bs1}, i. e. spline expansion
of the Nakanishi weight functions
\[ g_{i}(\gamma,z) = \sum_{\alpha\beta} g_{i,\alpha \beta}
S_{\alpha}(\gamma)S _{\beta}(z) Ê\]
on a compact domain $\Omega=[0,\gamma_{max}]\times[-1,+1]$ and
validation of the equations in some  well chosen points in the
$N_{\gamma}\times N_z$ corresponding intervals. The unknown
coefficients $g_{i,\alpha \beta}$ and the total mass $M$ are
obtained by solving a generalized eigenvalue problem
\[  \lambda N(M)g = A(M) g \]
As in the scalar case, the discretized left hand side kernel
$N$  has  very small eigenvalues
which make difficult the solution using standard methods. This
is avoided by adding a small constant term of the form
\[ N=N+ \epsilon_R I \]
where $I$ is the identity operator in the spline basis. The
error in the eigenvalues thus induced  is of the order of
$\epsilon_R$ and we have taken $\epsilon_R=10^{-6}$.

We have computed the binding energies,  defined as  $B=2m-M$,
and BS   amplitudes, for the $J=0^+$ two fermion system
interacting with  massive  scalar (S) and pseudoscalar (PS)
exchange kernels and for the fermion-antifermion system
interacting with massless vector exchange in Feynman gauge. In
the limit of an infinite vertex form factor parameter
$\Lambda\to\infty$, the later case would correspond to
positronium with an arbitrary value of the coupling constant.
All the results presented in this section are given in the
constituent mass units ($m=1$)   and with $L=1.1$.

\begin{figure}[ht!]
\vspace{.5cm}
\begin{center}\includegraphics[width=8cm]{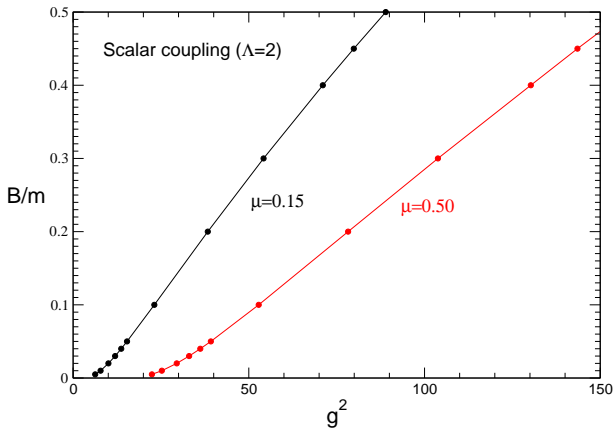}
\end{center}
\caption{Binding energy for scalar exchange v.s. $g^2$ for
$\Lambda=2$,  $L=1.1$, $\mu=0.15$ and $\mu=0.5$.}\label{Fig_B_g2_S}
\end{figure}

The binding energies obtained with the form factor parameter
$\Lambda=2$ are given in table \ref{tab_B_S_Ps}. For the scalar
and pseudoscalar cases,  we present the results for $\mu=0.15$
and $\mu=0.50$  boson masses. They have been compared to those
obtained in a previous calculation in Euclidean space
\cite{dorkin,Dorkin_PC} using a slightly different form factor.
Notice, that contrary to what is written in eq. (16) of
\cite{dorkin} -- which coincides with our form factor
(\ref{ffN}) -- the vertex form factor used in these calculations
is  \cite{Dorkin_PC}
\[  F(q^2)= {\Lambda^2\over\Lambda^2-q^2+i\epsilon} \]
Once taken into account this correction, our scalar  results are
in  full  agreement (four digits) with \cite{dorkin,Dorkin_PC}.
The pseudoscalar ones show small discrepancies ($\approx
0.5\%$). We have also computed \cite{CK_Euc} the binding
energies by directly solving the fermion BS equation the
Euclidean space using a method independent of the one used in
\cite{dorkin}. Our Euclidean results are in full agreement with
those given in the table \ref{tab_B_S_Ps}.

\begin{figure}[ht!]
\vspace{.5cm}
\begin{center}
\includegraphics[width=8cm]{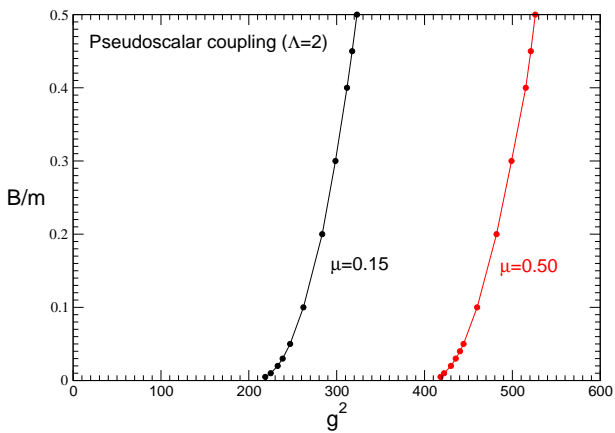}
\caption{Binding energy for pseudoscalar exchange v.s. $g^2$ for
$\Lambda=2$, $L=1.1$, $\mu=0.15$ and
$\mu=0.5$.}\label{Fig_B_g2_Ps}
\end{center}
\end{figure}

The $B(g^2)$ dependence for the scalar and pseudoscalar couplings
is plotted in figs. \ref{Fig_B_g2_S}  and \ref{Fig_B_g2_Ps}.
Notice the different $g^2$ scales of both dependences. The
pseudoscalar binding energies are fast increasing functions of
$g^2$ and thus more sensitive to the accuracy of numerical
methods. This sharp behaviour was also exhibit when solving the
corresponding Light-Front  equation \cite{MCK_PRC68_2003}.

It is worth noticing that the stability properties of the BS
$J^{\pi}=0^{+}$ solutions for the scalar coupling  are very
similar to  the Light-Front ones. In the latter case, we have
shown \cite{MCK_PRD64RC_2003,MCK_PRD64_2003}  the existence of a
critical coupling constant $g_c$ below which the system is stable
without vertex form factor while  for $g>g_c$, the system
"collapses", {\it i.e.} the spectrum is unbounded from below. The
numerical value was found to be $\alpha_c=3.72$
\cite{MCK_PRD64_2003}, which corresponds to  $g_c=6.84$.
Performing the same analysis than in our previous work -- eq. (71)
from  \cite{MCK_PRD64_2003}  -- we  found that for BS equation the
critical coupling constant is $g_c=2\pi$, in agreement with
\cite{dorkin}. The $10\%$ difference between the numerical values
of $g_c$ is apparently due to the different contents of the
intermediate states in the two approaches. The ladder BS equation
incorporates effectively the so-called stretch-boxes diagrams
which are not taken into account in the ladder LF results.

\bigskip
The positronium case deserves some comments. First we would like
to notice that in our formalism,  the  singularity of the
Coulomb-like kernels in terms of the momentum transfer
\mbox{$1/(k-k')^2$} is absent. This is a combined consequence of
the Nakanishi transform (\ref{bsint}) -- which allows to
integrate over $k'$ analytically in the right hand side of the
BS equation (\ref{bsf4}) -- and of the consecutive light front
projection integral. After this integration, the Coulomb
singularity  does not anymore manifest itself in the kernel.
This can be explicitly seen in the kernel of the Wick-Cutkosky
model obtained in eq. (22) of our previous work \cite{bs1}.

A second remark concerns the $\Lambda$ dependence of the
positronium  results. Using the methods developed in
\cite{MCK_PRD64RC_2003,MCK_PRD64_2003}  we found that in the BS
approach with ladder kernel there also exists  a critical value
of the coupling constant $g_c=\pi$. Note that, as in the scalar
coupling, the very existence and  the value of this critical
coupling constant is independent on the constituent ($m$) and
exchange  $\mu$ masses but depends on the quantum number of the
state.

\begin{table}[ht!]
\begin{center}
\caption{Coupling constant $g^2$ as a function of  binding energy
$B$   for  the positronium $J=0$ state in BS equation in the
region of stability without vertex form factor
($\Lambda\to\infty$), {\it i.e.} $g<\pi$. They are compared to the
non relativistic results. } \label{tab_B_Positronium}
\begin{tabular}{c|cc |cc |cc |c}
$B$    &$g^2_{NR}$&$g^2_{BS}$               \\ \hline
0.01   &   2.51          &    3.18        \\
0.02   &   3.55          &    4.65 \\
0.03   &   4.35          &    5.75   \\
0.04   &   5.03          &    6.64    \\
0.05   &   5.62          &    7.38   \\
0.06   &   7.95          &    8.02 \\
0.07   & 11.24          &   8.57 \\
0.08   & 13.77          &   9.06  \\
0.09   & 15.90          &   9.49\\
\end{tabular}
\end{center}
\end{table}

The ground state  positronium  binding energies without vertex
form factor are given in table  \ref{tab_B_Positronium}  for
values of the coupling below $g_c$, Nonrelativistic results
\[Ê g^2_{NR}=8\pi \sqrt{B/m} \]
are included for comparison. One can see that  the relativistic
effects  are  repulsive.

\begin{figure}[ht!]

\vspace{.5cm}
\begin{center} \includegraphics[width=8cm]{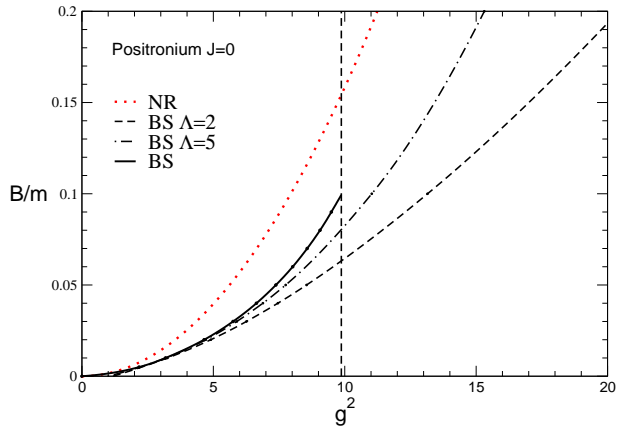}
\end{center}
\caption{Binding energy for $J=0$ positronium state versus $g^2$
(black solid line) in the stability region $g<g_c=\pi$. Dashed and
dotted-dashed curves correspond to the results for increasing
values of the vertex form factor parameter $\Lambda=2$ and
$\Lambda=5$ respectively. They are compared to the non
relativistic results (dotted line).} \label{Fig_B_g2_Positronium}
\end{figure}

These results  are displayed in fig. \ref{Fig_B_g2_Positronium}
(black solid line), and compared to the binding energies obtained
with two values of the form factor parameter $\Lambda=2$ (dashed)
and $\Lambda=5$ (dot-dashed). The stability region is limited by a
vertical dotted line  at $g=g_c=\pi$. Beyond this value the
binding energy without form factor becomes infinite and we have
found $B(g\to g_c)\approx0.10$.

\begin{figure}[ht!]
\vspace{.5cm}
\begin{center}
\includegraphics[width=7.8cm]{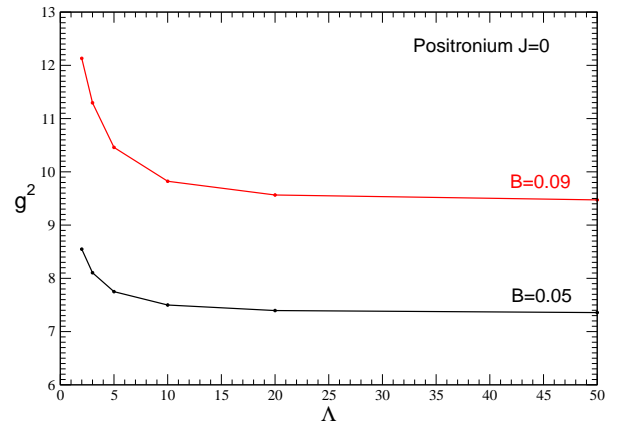}
\end{center}
\caption{$\Lambda$-dependence of $g^2$ for $J=0$ positronium state
for fixed values of binding energies $B=0.05$ and $B=0.09$.}
\label{Fig_g2_Lambda}
\end{figure}

The inclusion of the form form factor has a  repulsive effect,
{\it i.e.}  for a fixed value of the coupling constant it provides
a binding energy of the system which is smaller than in the
$\Lambda\to\infty$ limit (no cut-off). This is also illustrated in
fig. \ref{Fig_g2_Lambda} where we have plotted the $\Lambda$
dependence of $g^2$ for two different energies. One can see that
the value of the coupling constant to produce a bound state is a
decreasing function of $\Lambda$. The size of the effect depends
strongly on the binding energy but for both energies the
asymptotics is reached at $\Lambda\approx 20$. This behaviour is
understandable in terms of regularizing the short range
singularity of $-1/r^2$ interactions.

\begin{figure}[ht!]
\vspace{.5cm}
\begin{center}
\includegraphics[width=8cm]{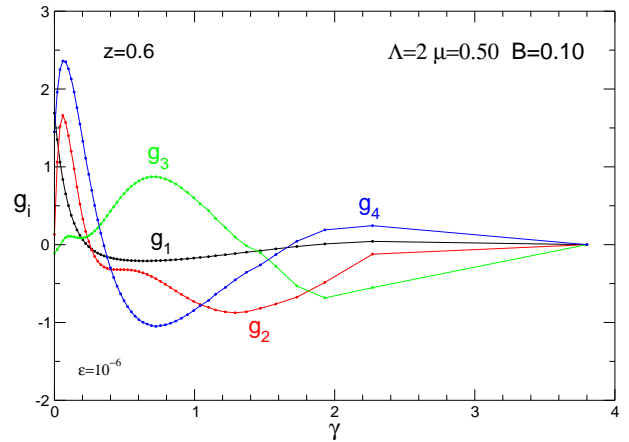}
\vspace*{1.cm}

\includegraphics[width=8cm]{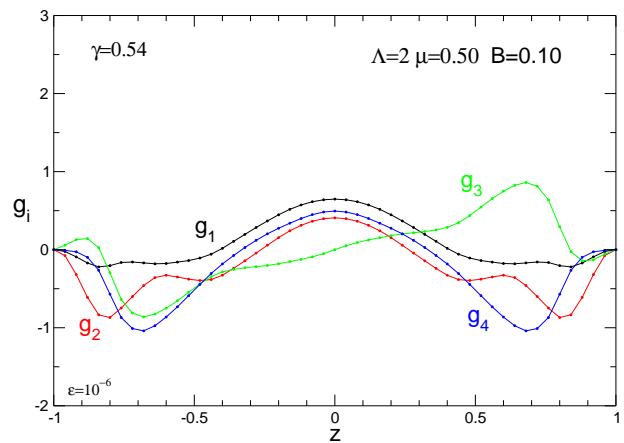}
\caption{Nakanishi weight functions for  scalar exchange for
$\Lambda=2$,  $B=0.10$, $\mu=0.50$ and $L=1.1$.}\label{gi}
\end{center}
\end{figure}

\bigskip
Finally we present some examples of the Nakanishi weigh functions
$g_i(\gamma,z)$. They correspond to a $B=0.1$ state with the
scalar coupling  and the same  parameters $\Lambda=2$, $\mu=0.50$
than in table \ref{tab_B_S_Ps}. They are displayed in fig.
\ref{gi}. In the upper figure is shown the $\gamma$-dependence for
a fixed value of $z$ and in the lower figure the $z$-dependence
for a fixed $\gamma$. Notice the regular behaviour of these
functions as well as their well defined parity with respect to $z$
-- $g_{1,2,4}$ are even and $g_3$ is odd --  consequence of
relations (\ref{phi_parity}) . As in the scalar case the
$\epsilon_R$-dependence of $g_i$ is more important than for the
binding energy.

\begin{figure}[ht!]
\begin{center}
\includegraphics[width=7.8cm]{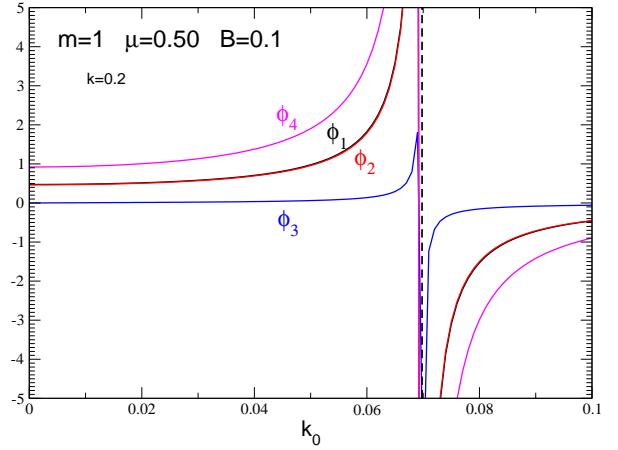}
\vspace*{1cm}

\includegraphics[width=7.8cm]{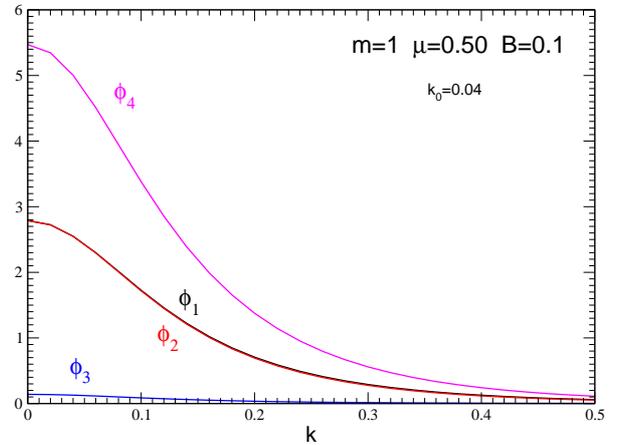}
\caption{Bethe-Salpeter Minkowski amplitudes   corresponding to
fig. \ref{gi}. Here $k=\mid\vec{k}\mid$. The amplitudes $\phi_1$
and $\phi_2$ are indistinguishable. }\label{Plot_Phi}
\end{center}
\end{figure}

Corresponding BS amplitudes $\phi_i$ are displayed in fig.
\ref{Plot_Phi}. Their computation  by  directly applying equation
(\ref{Nakanishi_phi})  is not very suitable due to the zeroes  of
the integrand denominator. We have preferred to compute $\phi_i$
by first factorizing the pole singularities in the right hand side
of  eq. (\ref{bsf4})
\[  \phi_a (k,p)     = \frac{i^2\tilde\Gamma_a(k,p)}{
{[(\frac{1}{2}p+k)^2-m^2+i\epsilon]}
{[(\frac{1}{2}p-k)^2-m^2+i\epsilon]} }  \] and computing
$\tilde\Gamma_a$ which is a regular function of $k$. By
inserting the Nakanishi transform (\ref{Nakanishi_phi}) in the
expression for $\tilde\Gamma_a$, the  integral over $k'$  in
(\ref{bsf4})  can be performed analytically and obtains the form
\[  {\tilde\Gamma_a}(k,p)= \sum_{a'} \int_0^{\infty}
\mbox{d}\gamma\int_{-1}^{+1} \mbox{d}z\;  I_{aa'}(k,p,\gamma,z)
g_{a'}(\gamma,z)  \]
where $I_{aa'}$ is a more appropriate function for computation
purposes.

The upper figure \ref{Plot_Phi} represents the $k_0$ dependence of
$\phi_i$ for a fixed value of  $\mid\vec{k}\mid=0.2$. They exhibit
a singular behaviour which corresponds to the pole of free
propagators $k_0=\epsilon_k-{M\over2}$. The lower figure
represents  the $\mid\vec{k}\mid$ dependence of  the amplitudes
$\phi_i$ for a fixed value  $k_0=0.04$. For this choice of
arguments, the amplitudes are  smooth functions of
$\mid\vec{k}\mid$, though they will be also singular for
$k_0>{B\over2}=0.05$. Notice that all the infinities in the BS
amplitudes $\phi_i$ come from the free propagator. However
$\tilde\Gamma_a$, although being  smooth, has its own
non-analytical points. For instance it obtains an imaginary part
for $({p\over2}\pm k)^2>(m+\mu)^2$ .

\section{Conclusions}\label{concl}

We present a new method for obtaining the solutions of the
Bethe-Salpeter equation in  Minkowski space for the two-fermion
system. It  is based on a Nakanishi integral representation of
the  amplitude and Light-Front projection and constitutes a
natural extension our previous work  for the scalar case
\cite{bs1}.

A straightforward generalization of this  approach, however
results into a singular fermionic kernel. In order to smooth this
singularities, a proper regularization of  the kernels has been
proposed. This generates a family of strictly equivalent equations
depending on one parameter $L$. Their solution gives the same
binding energies and Nakanishi weight functions $g_i(\gamma,z)$ .

The binding energies for the  scalar and pseudoscalar exchange
kernels  and for massless vector exchange (positronium)  are
found. They coincide with the ones found via Euclidean space
solution, thus providing a validity test of our method. The
solutions for the scalar and positronium states without vertex
form  factor ($\Lambda\to\infty$) are found to be stable below
some critical value $g_c$ of the coupling constant, respectively
$g_c=2\pi$ (scalar) and $g_c=\pi$ (positronium).

The BS amplitudes are obtained in terms of the computed
Nakanishi  weight functions. The exhibit a singular behaviour
due, on one hand, to the poles of the free propagators and, on
the other hand, to non analytic branch cuts which are
responsible for the appearance of an imaginary part.

An alternative approach to avoid the singularities of the
Minkowski amplitude would be to  solve the BS equation in the
Euclidean space in terms of  the Nakanishi weight functions.
Once solved, we can again restore the Minkowski BS amplitude by
the integral (\ref{bsint}). These different methods should
provide the same solutions although with  different precision
and stability. The results of this study will be published in a
future paper \cite{CK_Euc}.

\section*{Acknowledgements}
The authors are indebted to S.M.~Dorkin  and L.P.~Kaptari  for
sending detailed numerical results \cite{Dorkin_PC} of  their
work \cite{dorkin} and for many useful discussions. One of the
authors (V.A.K.) is sincerely grateful for the warm hospitality
of the theory group at the Laboratoire de Physique Subatomique
et Cosmologie, Universit\'e Joseph Fourier, in Gre\-no\-ble,
where part of this work was performed. Numerical calculations
were fulfilled at Centre de Calcul IN2P3 in Lyon.

\appendix

\section{The coefficients $C_{ij}$}\label{C_ij}

The spin-coupling matrix elements
$C_{ij}(\gamma,z;v)$ in eq. (\ref{W}) determining the kernel of
the equation (\ref{BSMF}), and also the kernel (\ref{W_ab}) at
$\nu=a$, for the $J=0$ state and scalar coupling are:
\[\begin{array}{lcl}
C_{11} &=&    \frac{1}{(1-z)} \; \frac{Q_+}{4m^2}   \\
C_{12} &=&    \frac{M}{4m} \\
C_{13} &=& 0 \\
C_{14} &=&  - {(1-v)\over(1-z)^2} \; { K_+ \; K_-\over 4m^2 M^2} \\
C_{21} &=&   C_{12}     \\
C_{22} &=&  -\frac{(\gamma+m^2)^2}{(1-z)^2 2m^2M^2}
+ \frac{\gamma z+(2-z)m^2}{4m^2(1-z)}  +  \frac{(1-z^2)M^2}{32m^2}   \\
C_{23} &=& - \frac{1-v}{(1-z)^3}\; {K_+ \;K_- \; Q_-\over 2m^2M^4} \\
C_{24} &=&    - {1-v\over (1-z)^2} \; {K_+\;K_-\over 2mM^3}   \\
C_{31} &=& 0     \\
C_{32} &=&        {1\over (1-z)} \; \frac{Q_-}{2m^2}\\
C_{33} &=&        {1-v\over(1-z)^2}  \;  {Q_- \; Q_+\over 2m^2M^2}\\
C_{34} &=&        {1-v\over 1-z} \; \frac{Q_-}{2mM}  \\
C_{41} &=&                   \frac{M^2}{4m^2} \\
C_{42} &=&  2\; C_{12}    \\
C_{43} &=&  C_{34}
              \\
C_{44} &=&  - {1-v\over 1-z} \frac{4\gamma - 4(1-2z)m^2 +
(1-z)^2M^2}{16m^2}
\end{array}\]
with
\begin{eqnarray*}
K_{\pm} &=&  \gamma + \left( m\pm {M\over2}(1-z) \right)^2\\
Q_{\pm} &=&  \gamma + m^2 \pm {M^2\over4}(1-z)^2     
\end{eqnarray*}

The coefficients for the pseudoscalar and massless vector exchange
are given by the relations (\ref{C_Ps})  and (\ref{C_V})
respectively.

\section{Derivation of the equation (\ref{eq0f})}\label{Derivation_Vd}

In order to derive the equation (\ref{eq0f}), we substitute in
eq. (\ref{bsf4p}) the amplitudes $\phi_i(k,p)$ in the form
(\ref{bsint}), shift the variable $k\to
k+\frac{\omega}{\omega\cd p}\beta$ and integrate over $\beta$ in the
infinite limits.

The  integrand depends on the scalar variables $k^2,\;p\cd k$ and
$\frac{\omega\cd k}{\omega\cd p}$ which are not independent but
are related to each other by a constrain. Following appendix A
from \cite{bs1}, we parametrize them in terms of two independent
variables $\gamma,z$ as:
\begin{eqnarray*}
k^2&=& -\frac{(\gamma+z^2 m^2)}{1-z^2} \\
\frac{\omega\cd k}{\omega\cd p} &=&-\frac{1}{2}z \\
p\cd k &=& \frac{z[\gamma+z^2 m^2+(1-z^2)\kappa^2]}{1-z^2},
\end{eqnarray*}
where $\kappa^2=m^2-\frac{1}{4}M^2$. In the spinless case
\cite{bs1}, when we applied integration over $\beta$ to the equation
without the factor $\eta(k,p)$, we found that its left-hand side
obtained the form:
\begin{eqnarray}\label{lhs2}
&&l.h.s.=
\int_{-1}^1\mbox{d}z'\int_0^{\infty}\mbox{d}\gamma'\int_{-\infty}^{\infty}d\beta
\\
&&\times\frac{g(\gamma',z')}{\left[\beta (z-z')+\gamma' +(1-z
z')\left(\kappa^2 + \frac{\gamma+m^2 z^2}{1-z^2}\right)-
i\epsilon \right]^3} \nonumber
\end{eqnarray}
As a function of $\beta$, it contains one pole at
\begin{equation}\label{beta0}
\beta'_0= \frac{+i\epsilon}{z-z'}.
\end{equation}
Taking the residue, we found that the result of integration is
proportional to $\delta(z'-z)$ and consequently the left-hand side obtained the
form of a one-dimensional integral:
$$
l.h.s.
=\int_0^{\infty}\frac{i\pi\,g(\gamma',z)\mbox{d}\gamma'}{\Bigl[\gamma'+\gamma
+z^2 m^2+(1-z^2)\kappa^2\Bigr]^2}
$$

The factor $\eta(k,p)$ itself consists of two factors. Therefore,
after replacement $k\to k+\frac{\omega}{\omega\cd p}\,\beta$,
the factor $\eta(k+\frac{\omega}{\omega\cd p}\beta,p)$ v.s.
$\beta$ obtains two poles:
$$
\beta_1=\frac{-i\epsilon}{1-z}+\ldots,\quad
\beta_2=\frac{+i\epsilon}{1+z}+\ldots,
$$
Integrating then the left-hand side of (\ref{bsf4p}) (with shifted
argument $k+\frac{\omega}{\omega\cd p}\beta$) over $\beta$, we can
take into account one of these poles, which is in the opposite
half-plane than the pole (\ref{beta0}). The position of the pole
(\ref{beta0}) depends on the sign of the difference $z-z'$. The
integration, after cancellation of both sides of equation by a
common factor, results into the left-hand side of eq.
(\ref{eq0f}). Now the delta-function \mbox{$\delta(z'-z)$} does
not appear and the integral is two-dimensional one.

Concerning the calculation of the right-hand side, the difference
relative to what was explained in \cite{bs1} for the spinless
particles is minimal. In the spinless case, the right-hand side
 already had two propagators, like in eq. (\ref{bsf4}). Their
poles v.s. $\beta$ (as well as the pole of the kernel) were
taken into account. The factor $\eta(k,p)$ adds two new
poles in other points, which differ from the scalar case by the
replacement $m\to L$ and which can be taken into account
in a similar way. After shifting the variable $k\to
k+\frac{\omega}{\omega\cd p}\,\beta$ the coefficients $c_{ij}$
in right-hand side of (\ref{bsf4}) become depend on $\beta$, but
they have no poles.

In this way, multiplying both parts of eq. (\ref{bsf4}) by
$\eta(k,p)$, integrating analytically over $k'$), shifting the
four-vector  $k\to k+\frac{\omega}{\omega\cd p}\,\beta$  and
integrating over $\beta$, we find the system of equations
(\ref{eq0f}).

\section{The coefficients $C^{b}_{ij}$}\label{Cb_ij}
The spin-coupling matrix $C^b_{ij}(\gamma,z;v)$ in the kernel
(\ref{W_ab}) at $\nu=b$ for the $J=0$ state and scalar coupling
are:
\[\begin{array}{lclcl}
C^b_{11} &=& C^a_{11}  &+&   {z(L^2-m^2)\over 4m^2(1-z)}     \\
C^b_{12} &=& C^a_{12}  &  &                           \\
C^b_{13} &=& 0                 &  &     \\
C^b_{14}&=&                     & -&  {1-v\over (1-z)^2} \; {K^b_+ K^b_-\over 4m^2M^2}\\
C^b_{21} &=&  C^a_{21}  &  &  \\
C^b_{22} &=&  C^a_{22}  &-&     \frac{   (L^2-m^2) \;  [4\gamma + 2m^2  + 2L^2 -(1-z)M^2  ]   }{4(1-z)^2m^2M^2}  \\
C^b_{23} &=&                     &-&  \frac{1-v}{(1-z)^3}\; {K^b_+ K^b_- Q^b_-\over 2m^2M^4} \\
C^b_{24} &=&                     &-&  {1-v\over (1-z)^2} \; {K^b_+\;K^b_-\over 2mM^3}   \\
C^b_{31} &=& 0                 & &  \\
C^b_{32} &=&                    & & {1\over (1-z)} \; \frac{Q^b_-}{2m^2}   \\
C^b_{33} &=&  C^a_{33} &+&     {1-v\over(1-z)^2} \frac{(L^2-m^2)
\;  [4\gamma
 + 2m^2  + 2L^2 -(1-z)M^2  ]   }{4m^2M^2}  \\
C^b_{34} &=&                    & &  {1-v\over1-z}\; \frac{Q^b_- }{2mM} \\
C^b_{41} &=& C^a_{41}  &  &  \\
C^b_{42} &=& 2C^b_{12} &=& C^a_{42}      \\
C^b_{43} &=& C^b_{34}
\\
C^b_{44} &=& C^a_{44}  &- &  {(1-v)z(L^2-m^2)\over 4m^2(1-z)}
\end{array}\]
with
\begin{eqnarray*}
K^b_{\pm} &=&  \gamma + \left( L\pm {M\over2}(1-z) \right)^2\\
Q^b_-  &=&  \gamma + L^2 - {M^2\over4}(1-z^2)
\end{eqnarray*}
and $C^a_{ij}  =C_{ij} $ given in Appendix  \ref{C_ij}.

The relations with the pseudoscalar  and massless vector exchange
are  also given by (\ref{C_Ps})  and (\ref{C_V}) respectively.



\begin{thebibliography}{10}

\bibitem{bs1}
V.A.~Karmanov and J.~Carbonell,
Eur. Phys. J. A {\bf 27}, 1 (2006); hep-th/0505261.


\bibitem{BS}
E.E.~Salpeter and H.A.~Bethe, Phys. Rev. {\bf 84}, 1232 (1951).

\bibitem{Maris1}
   P.~Maris and P.~C.~Tandy,
   Nucl.\ Phys.\ B\ (Proc.\ Suppl.)\  {\bf 161} 136 (2006);
   [arXiv:nucl-th/0511017].

\bibitem{Maris2}
   M.~S.~Bhagwat and P.~Maris,
   Phys.\ Rev.\  C {\bf 77} 025203 (2008); [arXiv:nucl-th/0612069].

\bibitem{ckm_ejpa}
J.~Carbonell, V.A.~Karmanov, M.~Mangin-Brinet,
Eur. Phys. J. A {\bf 39},  53 (2009); arXiv:0809.3678 (hep-th).

\bibitem{KW} K.~Kusaka, A.G.~Williams, Phys. Rev. D {\bf 51}, 7026 (1995).
\bibitem{KW1} K.~Kusaka, K.~Simpson, A.G.~Williams, Phys. Rev. D {\bf 56}, 5071 (1997).

\bibitem{SA_PRD67_2003} V.~Sauli, J~Adam, Jr., Phys. Rev. D {\bf 67}, 085007 (2003).

\bibitem{sauli}  V.~Sauli, J. Phys. G {\bf 35}, 035005 (2008); arXiv:0802.2955 [hep-ph].

\bibitem{bbmst} S.G.~Bondarenko, V.V.~Burov, A.M.~Molochkov,
G.I.~Smirnov and H.~Toki, Prog. in Part. and Nucl. Phys., {\bf
48}, 449 (2002).

\bibitem{bbpr}
S.G.~Bondarenko, V.V.~Burov, W.-Y.~Pauchy Hwang, E.P.~Rogochaya,
Nucl. Phys. A {\bf 832}, 233 (2010); arXiv:0810.4470 (nucl-th);
arXiv:1002.0487 (nucl-th).

\bibitem{nakanishi1}
N.~Nakanishi, Phys. Rev.  {\bf 130}, 1230 (1963); Prog. Theor.
Phys. Suppl. {\bf 43}, 1 (1969).

\bibitem{nakanishi2}
N.~Nakanishi, {\it Graph Theory and Feynman Integrals}, Gordon and
Breach, New York, 1971.

\bibitem{bs2}
J.~Carbonell and V.A.~Karmanov,
Eur. Phys. J. {\bf A27}, 11 (2006); hep-th/0505262.

\bibitem{dorkin} S.M.~Dorkin, M.~Beyer, S.S.~Semykh and
L.P.~Kaptari, Few-Body Syst. {\bf 42}, 1 (2008); arXiv:0708.2146
(nucl-th).

\bibitem{Dorkin_PC} S.M.~Dorkin and L.P.~Kaptari, private communication.

\bibitem{CK_Euc}  J.~Carbonell and V.A.~Karmanov, to be published.

\bibitem{MCK_PRC68_2003} M. Mangin-Brinet, J. Carbonell, V. Karmanov, Phys. Rev. C {\bf 68}, 055203 (2003).

\bibitem{MCK_PRD64RC_2003} M. Mangin-Brinet, J. Carbonell, V. Karmanov, Phys. Rev. D {\bf 64}, 027701 (2001).

\bibitem{MCK_PRD64_2003}  M. Mangin-Brinet, J. Carbonell, V. Karmanov,  Phys. Rev. D {\bf 64}, 125005 (2001).




\end{thebibliography}
\end{document}